\documentclass[iicol]{sn-jnl}

\jyear{2022}%

\theoremstyle{thmstyleone}%
\theoremstyle{thmstyletwo}%

\theoremstyle{thmstylethree}%

\usepackage[utf8]{inputenc}

\usepackage[utf8]{inputenc}
\usepackage[T1]{fontenc}
\usepackage[OT1]{fontenc}
\usepackage{mathrsfs}
\usepackage{graphicx}
\usepackage{amssymb}
\usepackage{amsbsy}
\usepackage{latexsym}
\usepackage{mathtools}
\usepackage{titlesec}
\usepackage{amsmath}
\usepackage{color}
\usepackage[utf8]{inputenc}
\usepackage{lipsum}
\usepackage{fontenc}
\usepackage{dcolumn}
\usepackage{bbold}
\usepackage{slashed}
\usepackage{cite}

\usepackage{chngcntr}
\numberwithin{equation}{section}
\renewcommand{\theequation}{\arabic{section}.\arabic{equation}}

\def\nn{\nonumber}

\def\b{\beta}
\def\g{\gamma}

\def\m{\mu}
\def\n{\nu}
\def\r{\rho}
\def\s{\sigma}

\def\ds{\stackrel{\star}{,}}
\def\de{\mathrm{d}}
\def\De{\textrm{D}}

\def\tr{{\rm Tr}}

\def\dj{d\kern-0.4em\char"16\kern-0.1em}
\def\Dj{\mbox{\raise0.3ex\hbox{-}\kern-0.4em D}}

\begin{document}

\title[0]{Noncommutative $SO(2,3)_{\star}$ Gauge Theory of Gravity}



\author*[1]{\fnm{Marija} \sur{Dimitrijevi\'{c} \'{C}iri\'{c}}}\email{dmarija@ipb.ac.rs}

\author[1]{\fnm{Du\v{s}an} \sur{\Dj or\dj evi\'{c}}}\email{djdusan@ipb.ac.rs}

\author[1]{\fnm{Dragoljub} \sur{Go\v{c}anin}}\email{dgocanin@ipb.ac.rs}

\author[1]{\fnm{Biljana} \sur{Nikoli\'{c}}}\email{biljana@ipb.ac.rs}

\author[1]{\fnm{Voja} \sur{Radovanovi\'{c}}}\email{rvoja@ipb.ac.rs}

\affil[1]{\orgdiv{Faculty of Physics}, \orgname{University of Belgrade}, \orgaddress{\street{Studentski Trg 12-16}, \city{Belgrade}, \postcode{11000}, \country{Serbia}}}

\abstract{Topological gravity (in the sense that it is metric-independent) in a $2n$-dimensional spacetime can be formulated as a gauge field theory for the AdS gauge group $SO(2,2n-1)$ by adding a multiplet of scalar fields. These scalars can break the gauge invariance of the topological gravity action, thus making a connection with Einstein's gravity. This review is about a noncommutative (NC) star-product deformation of the four-dimensional AdS gauge theory of gravity, including Dirac spinors and the Yang-Mills field. In general, NC actions can be expanded in powers of the canonical noncommutativity parameter $\theta$ using the Seiberg-Witten map. The leading-order term of the expansion is the classical action, while the higher-order $\theta$-dependent terms are interpreted as new types of coupling between classical fields due to spacetime noncommutativity. We study how these perturbative NC corrections affect the field equations of motion and derive some phenomenological consequences, such as NC-deformed Landau levels of an electron. Finally, we discuss how topological gravity in four dimensions (both classical and noncommutative) appears as a low-energy sector of five-dimensional Chern-Simons gauge theory in the sense of Kaluza-Klein reduction.}



\maketitle

\section{Introduction}\label{sec1}

The Standard Model (SM) of particle physics is considered to be one of the most successful theoretical constructions in modern physics. It is based on the principle of gauge invariance and formulated as a Yang-Mills theory for a suitable gauge group. Despite its success in describing other fundamental interactions, the SM does not include gravity. According to Einstein's theory of General Relativity (GR), gravity is a manifestation of the curvature of spacetime, and the whole metric formulation of the theory is given in terms of pseudo-Riemannian geometry. In search of unification, many attempts have been made to develop a proper gauge theory of gravity. 
Since the original papers of Utiyama \cite{Utiyama}, Kibble \cite{Kibble} and Sciama \cite{Sciama}, there has been considerable interest in this subject; instead of giving a complete historical account, we refer to several available reviews \cite{Rev, Rev2, Rev3}. The main result of these efforts has been to establish a connection between GR, expressed in the first-order formalism, and the Poincar\'{e} gauge theory, with the spin-connection representing a gauge field for local Lorentz rotations and the vierbein field being considered as a gauge field for spacetime translations \cite{PGT1, PGT2}. However, the analogy with the Yang-Mills gauge theory is incomplete because of the specific treatment of translations, which are much harder to incorporate. Their nonlinear realization is the reason why this theory is not a proper gauge theory, in the sense that it is not founded on a fibre bundle structure. 

One particular approach, and the one that we will put our focus on in this review, is developed in the works of MacDowell-Mansouri \cite{MacMan}, Stelle and West \cite{stelle-west, west}, Wilczek \cite{Wilczek}, and Chamseddine and Mukhanov \cite{Mukhanov1, Mukhanov2}. The main idea behind this approach is to unify the spin-connection and the vierbein field as components of a single gauge field of some larger gauge group, in particular, the AdS group $SO(2,3)$, that contains the Lorentz group as a subgroup. For a more mathematical viewpoint concerning the geometrical basis of the MacDowell-Mansouri gravity, see \cite{Wise}. A major motivation for studying gauge-theoretic formulations of gravity based on the $SO(2,3)$ gauge group (instead of $SO(1,4)$, for example) came with the development of Supergravity (SUGRA). Namely, the AdS algebra $\mathfrak{so}(2,3)$ is the bosonic sector of the orthosymplectic superalgebra $\mathfrak{osp}(4\vert 1)$, which is used to define a generalization of SUGRA to include the cosmological constant, see \cite{MacMan, west, Townsend, Ferrara, Preitschopf} and for a comprehensive treatise on the subject \cite{Ortin}. 

The action originally introduced by MacDowell and Mansouri, and subsequently studied extensively by other authors, was formulated entirely using objects that transform under the $SO(2,3)$ gauge group, but with a term that breaks the symmetry directly in the Lagrangian. We will consider a slightly modified version, introduced by Stelle and West, that at the first stage ensures manifest $SO(2,3)$ invariance,
\begin{equation} \label{topgravnep}
\int\varepsilon_{ABCDE}F^{AB}\wedge F^{CD}\phi^{E}, \end{equation}
where we have a field strength $2$-form  $F^{AB}$.
This action is not of the Yang-Mills type, but it is topological in the sense that the Lagrangian does not involve the metric tensor \footnote{However, this fact does not imply that there are no propagating local degrees of freedom, and thorough consideration shows that such degrees of freedom do exist \cite{Morales:2016qrx}.}.
The action (\ref{topgravnep}) was also studied by Chamseddine in \cite{Chamseddine_topo} as a $D=4$ case of a wider class of topological gravity theories defined in all even dimensions. For example, in two dimensions, the action is given by
\begin{equation}
\int \hspace{1mm}\varepsilon_{ABC}F^{AB}\phi^{C},
\end{equation}
and it corresponds to a general class of topological BF theories that can reproduce \textit{JT}-gravity model for a suitable gauge group, with an on-shell vanishing torsion \cite{Chamseddine:1989wn, Isler:1989hq}. For phenomenological reasons, we will stick to the case of $D=4$. However, as explained in \cite{Chamseddine_topo}, this kind of topological gravity in even-dimensional spacetime demands an introduction of a multiplet of spacetime scalar fields $\phi^A$ that transform as a vector field with respect to the gauge group. Furthermore, in MacDowell-Mansouri-like models, it is constrained using a gauge invariant condition $\phi^A\phi_A=l^2$, where $l$ stands for the AdS radius. Stelle and West introduced this condition using a Lagrange multiplier, though we will not tackle this approach in the further text. 

One must note, however, that there is no experimental evidence for an enhanced local symmetry of a gravitational theory, which seems to indicate that one must break $SO(2,3)$ gauge symmetry down to its Lorentz $SO(1,3)$ subgroup. One can break the symmetry directly by setting $\phi^A=l\delta_5^A$; in this way, the local Lorentz symmetry will be preserved. This symmetry breaking results in a Lagrangian for gravity that consists of the Einstein-Hilbert term with a cosmological constant and the topological Gauss-Bonnet term. Scalar field $\phi$ can be therefore considered as a background field whose equations of motion we don't consider. 

While it is necessary to introduce a scalar field to define the action (\ref{topgravnep}), the physical motivation remains obscure. For example, topological gravity action can be constructed for any odd-dimensional manifold using an appropriate Chern-Simons form. The resulting theory is a gauge theory for a Chern-Simons connection built from the frame field and the spin-connection without introducing any other constituent field. It was shown in \cite{Chamseddine_topo} that dimensional reduction of the $(2n+1)$-dimensional Chern-Simons action leads directly to a $(2n)$-dimensional topological action (\ref{topgravnep}). This construction suggests how scalar field could naturally arise in even-dimensional theory, at least classically.  

 
GR is power-counting nonrenormalizable; therefore, we cannot make any firm conclusion about the quantum gravity regime. A way to tackle questions concerning quantum aspects of spacetime is to assume that (on some energy scale) a valid consideration is given in terms of noncommutative (NC) geometry. This approach is based on the idea that the fundamental structure of spacetime is not captured by classical smooth manifolds, as in GR. Instead, one introduces an abstract algebra of non-commuting coordinates $\hat{x}^{\mu}$ satisfying some non-trivial commutation relations that describe a deviation from the classical structure. A particular implementation of this idea is to, instead of using non-commuting coordinates, introduce a deformation of the algebra of functions on ordinary spacetime (parametrized by ordinary commuting coordinates) in terms of an NC star-product. Conditions under which such a product can be consistently defined are given by the Poincar\'{e}-Birkhof-Witt (PBW) theorem \cite{NC_book}. The canonical situation is the so-called $\theta$-constant noncommutativity,
\begin{equation}
[\hat{x}^{\mu},\hat{x}^{\nu}]=i\theta^{\mu\nu},  
\end{equation} 
where $\theta^{\mu\nu}$ is an antisymmetric matrix of constant deformation parameters. This case of NC geometry corresponds to the Moyal-Groenewold-Weyl star-product, defined by
\begin{align}\label{Moyal}
(f\star g)(x)&=f(x)e^{\frac{i}{2}\overleftarrow{\partial}_{\mu}\theta^{\mu\nu}\overrightarrow{\partial}_{\nu}}g(x)
\\
&=f(x)g(x)+\frac{i}{2}\theta^{\mu\nu}\partial_{\mu}f(x)\partial_{\nu}g(x)+\dots.\nonumber
\end{align}
The leading-order term in the $\theta$-expansion is the ordinary commutative point-wise product of functions, while the higher-order $\theta$-dependent terms represent NC corrections. See \cite{NC_book, Leo_NC, Szabo} for a comprehensive review of the subject.

The Seiberg-Witten (SW) map originally introduced in \cite{SW} is a valuable tool in studying NC gauge field theories. It allows us to define an NC gauge field theory by organizing NC fields in terms of their commutative counterparts in a way that preserves the number of degrees of freedom of the classical theory and consistently generalizes classical symmetries. We can use the SW map to expand NC actions in powers of $\theta$ and obtain an equivalent commutative theory with $\theta$-dependent terms interpreted as new couplings between classical fields that ought to capture some underlying quantum aspects of spacetime. 


NC gravity has been studied extensively for the past twenty years from various viewpoints. In \cite{SWmapApproach1, SWmapApproach2, SWmapApproach3} an NC deformation of pure Einstein gravity based on the SW construction is proposed. The twist approach \cite{TwistApproach1, TwistApproach2} was used to analyze some particular NC solutions 
\cite{TwistSolutions1, TwistSolutions2}. Using twisted differential geometry tools, a geometric theory of NC gravity was constructed in \cite{PLM-13} and extended to include fermions and gauge fields in \cite{PLGR-fer1, PLGR-fer2, Paolo-fer, PLgaugefield}. The concept of Lorentz symmetry in NC gauge field theories was further considered in \cite {Chaichian1, Chaichian2}. In the case of emergent NC gravity, dynamical quantum geometry arises from an NC gauge 
theory given by the Yang-Mills
matrix models \cite{EmGravityApproach1, EmGravityApproach2}. There are also fuzzy space gravity models \cite{Fuzzy1, Fuzzy2}. On a more mathematical side, the SW map was related to NC gravity 
models via the Fedosov deformation quantization of endomorphism bundles 
\cite{Dobrski1, Dobrski2}. 
Other attempts to relate NC gravity models with some 
testable GR results like gravitational waves, cosmological solutions and
Newtonian potential can be found in \cite{OtherApproaches1, OtherApproaches2, OtherApproaches3, OtherApproaches4, OtherApproaches5, OtherApproaches6}. 
A connection to SUGRA was established in \cite{LeoNCSUGRA, NCSUGRA1, NCSUGRA2} and the extension of NC gauge theories to orthogonal and symplectic algebras was considered in \cite{NCsymplectic1, NCsymplectic2}. 

Finally, a canonical NC deformation of a gauge theory of gravity based on the action (\ref{topgravnep}) and AdS gauge group $SO(2,3)$ was developed in \cite{Us1, Us2, Us3, Us4}. The first non-vanishing NC correction to GR action is confirmed to be of the second-order. This result is in accord with \cite{PLM-13}. The $SO(2,3)_{\star}$ model of NC gravity can also incorporate matter fields \cite{VG, U(1), YM} and has a natural extension to NC SUGRA \cite{VG-SUGRA}.

The purpose of this review is to present some of the main results obtained by applying the tools of NC field theory to the classical AdS gauge theory of gravity and comparing them with the related results in the literature. In Section 2, we set up the classical AdS gauge theory of gravity, including Dirac spinors and the Yang-Mills field. An NC extension of this classical model is developed in Section 3, with a special emphasis on using the Seiberg-Witten map. In Section 4, we consider a special case of NC Electrodynamics and analyze how noncommutativity affects Landau levels of an electron. Finally, in Section 5, we discuss how the four-dimensional AdS gauge theory of gravity can be understood starting from the five-dimensional Chern-Simons gauge theory through Kaluza-Klein reduction.


\section{AdS gauge theory of gravity}

In this section, we introduce a classical (i.e. commutative) model of $D=4$ gravity based on a broken AdS gauge group $SO(2,3)$, and also include Dirac spinors and the Yang-Mills field.   
\subsection{Notation and conventions}

Four-dimensional AdS space, $AdS_{4}$, is a maximally symmetric space with Lorentzian signature $(+---)$ and constant negative curvature; it can be represented as a hyperboloid embedded in a five-dimensional flat ambient space with signature $(+---+)$. Lie group $SO(2,3)$ is the isometry group of $AdS_{4}$ and the corresponding Lie algebra $\mathfrak{so}(2,3)$ is spanned by ten generators, $M_{AB}=-M_{BA}$ ($A,B=0,1,2,3,5$), satisfying the AdS algebra relations
\begin{align}\label{AdSalgebra}
[M_{AB},M_{CD}]=i(&\eta_{AD}M_{BC}+\eta_{BC}M_{AD}\nn\\
-&\eta_{AC}M_{BD}-\eta_{BD}M_{AC}).
\end{align}
By splitting the generators into six AdS rotations $M_{ab}$ ($a,b=0,1,2,3$) and four AdS translations $M_{a5}$, we can recast the AdS algebra (\ref{AdSalgebra}) in a more explicit form,
\begin{align}
[M_{ab}, M_{cd}]&=i(\eta_{ad}M_{bc}+\eta_{bc}M_{ad}-(c\leftrightarrow d)) ,\nn\\
[M_{ab}, M_{c5}]&=i(\eta_{bc}M_{a5}-\eta_{ac}M_{b5}), \nn \\
[M_{a5}, M_{b5}]&=-iM_{ab}. \label{AdS_PP}
\end{align}
A realization of the AdS algebra (\ref{AdS_PP}) is provided by $5$D gamma-matrices $\Gamma^A$ satisfying Clifford algebra relations, $\{\Gamma_A,\Gamma_B\}=2\eta_{AB}\mathbb{1}_{4\times 4}$; the $SO(2,3)$ generators are given by $M_{AB}=\frac{i}{4}[\Gamma_A,\Gamma_B]$. One choice of the $5$D gamma matrices is $\Gamma_A =(i\gamma_a\gamma_5, \gamma_5)$, where $\gamma_a$ are the usual $4$D Dirac gamma matrices, and $\gamma_5=-i\gamma_0\gamma_1\gamma_2\gamma_3$. In this particular representation we have $M_{ab}=\frac{i}{4}[\gamma_a,\gamma_b]=\frac12\sigma_{ab}$ and $M_{5a}=\frac{1}{2}\gamma_a$. 

\subsection{Physical Theory}

We first introduce some basic building blocks of the AdS gauge theory of gravity. The $\mathfrak{so}(2,3)$-valued gauge field $\omega_{\mu}$ can be separated into two parts,
\begin{equation}\label{gauge_pot_AdS}
\omega_\mu = \frac{1}{2}\omega_\mu^{AB}M_{AB}=\frac{1}{4}\omega_\mu^{ab}\sigma_{ab}-
\frac{1}{2}\omega_\mu^{a5}\gamma_a.
\end{equation}
The corresponding field strength is defined in the usual way, 
\begin{equation}\label{FAB}
F_{\mu\nu}=\partial_\mu\omega_\nu-\partial_\nu\omega_\mu-i[\omega_\mu,\omega_\nu], 
\end{equation}
and it can also be separated according to the generators as
\begin{align}\label{Fab+Fa5}
F_{\mu\nu}=\frac{1}{4}\Big( R_{\mu\nu}^{ab}&-(\omega_\mu^{a5}\omega_\nu^{b5}-\omega_\nu^{a5}\omega_\mu^{b5})\Big)
\sigma_{ab}\nn\\ &-\frac{1}{2}F_{\mu\nu}^{a5}\gamma_a, 
\end{align}
where we have
\begin{align}\label{Rab}
R_{\mu\nu}^{ab} &= \partial_\mu\omega_\nu^{ab}+\omega_{\mu \;\;c}^{\;\;a}\;\omega_\nu^{cb}-(\mu\leftrightarrow\nu), \\
F_{\mu\nu}^{a5} &= D^{L}_\mu \omega^{a5}_\nu-D^{L}_\nu \omega^{a5}_\mu.\label{Ta}
\end{align}
Note that $D^{L}_{\m}$ stands for the Lorentz $SO(1,3)$ covariant derivative that corresponds to the $\omega^{ab}_{\mu}$ component of the gauge field.

Under infinitesimal $SO(2,3)$ gauge transformations, field strength transforms in the adjoint representation,
\begin{equation}
\delta_\epsilon F_{\mu\nu}=i[\epsilon, F_{\mu\nu}], \label{TrLawFAB}
\end{equation}
where $\epsilon$ stands for an $\mathfrak{so}(2,3)$-valued gauge parameter. From the transformation law (\ref{TrLawFAB}) follows that by setting $\epsilon^{a5}=0$ (by doing this we restrict the group of gauge transformations to $SO(1,3)$) we may identify $\omega^{ab}_\mu$ component of the AdS gauge field with the Lorentz $SO(1,3)$ spin-connection of, $\omega^{a5}_\mu$ with the (rescaled) vierbein $e_{\m}^{a}/l$, field strength component $R^{ab}_{\mu\nu}$ with the curvature tensor, and $F^{a5}_{\mu\nu}$ with (rescaled) torsion $T_{\m\n}^{a}/l$; the constant parameter $l$ has to be introduced on the ground of dimensional analysis, and as will become clear later, stands for the AdS radius.  

We start with the $SO(2,3)$-invariant action (\ref{topgravnep}) that involves an auxiliary non-dynamical background field $\phi=\phi^{A}\Gamma_{A}$. This field is a spacetime scalar and an internal space vector transforming in the adjoint representation of $SO(2,3)$, i.e. $\delta_{\epsilon}\phi=i[\epsilon,\phi]$.
It has length dimension $1$ and it is constrained by a gauge invariant condition $\phi^{2}=\eta_{AB}\phi^{A}\phi^{B}=l^{2}$. Using this field, one can define two additional $SO(2,3)$-invariant actions \cite{Wilczek} that will, together with (\ref{topgravnep}), constitute our AdS gauge theory of gravity,

\begin{align}\label{KomDejstvo_S_1}
S_1=\frac{il}{64\pi G_N}\int{\rm d}^4x\;\varepsilon^{\mu\nu\rho\sigma}\tr\left( 
F_{\mu\nu} F_{\rho\sigma}\phi\right),
\end{align}
\begin{align}\label{KomDejstvo_S_2}
S_2=\frac{1}{128 \pi G_{N}l}&\int \de^{4}x\;\varepsilon^{\mu \nu \rho 
\sigma} \\
&\times\tr\left(F_{\mu 
\nu}D_{\rho}\phi D_{\sigma}\phi\phi\right)+c.c., \nonumber
\end{align}
\begin{align}\label{KomDejstvo_S_3}
S_3=-\frac{i}{128 \pi G_{N}l}&\int \de^{4}x\;\varepsilon^{\mu \nu \rho 
\sigma}\\
&\times\tr\left(D_{\mu}\phi D_{\nu}\phi D_{\rho}\phi D_{\sigma}\phi\phi\right),\nonumber
\end{align}
with $SO(2,3)$ covariant derivative
\begin{equation}\label{Cov-phi}
D_{\mu}\phi=\partial_{\mu}\phi-i[\omega_{\mu},\phi].
\end{equation}
The complete commutative model of AdS gauge theory of gravity is defined by the sum of these three action terms,
\begin{equation}
S=c_1S_1+c_2S_2+c_3S_3, \label{FullCommAction}
\end{equation}
where we introduced free parameters $c_1,c_2$ and $c_3$ that will be determined by imposing some additional constraints. 

The symmetry is broken directly from $SO(2,3)$ to $SO(1,3)$ by fixing the value of the background field to $\phi^{A}=(0,0,0,0,l)$, i.e. $\phi=l\gamma_{5}$. This choice is consistent with the constraint $\phi_A\phi^A=l^2$ and it can be understood as a gauge choice.
In this ``physical gauge'', the classical action (\ref{FullCommAction}) becomes 
\begin{align}\label{KomDejstvo} 
&S=-\frac{1}{16\pi G_{N}}\int 
\de^{4}x\Bigg(\frac{c_1l^2}{16}\varepsilon^{\mu\nu\rho\sigma}R_{\mu\nu}^{mn}R_{\rho\sigma}^{rs}\varepsilon_{mnrs}\nn\\
+&\sqrt{-g}\Big((c_1 + c_2)R-\c{6}{l^2}(c_1+2c_2+2c_3)\Big)
\Bigg). 
\end{align}
By imposing the constraint $c_{1}+c_{2}=1$ the action reduces to the Einstein-Hilbert action with the cosmological constant given by
\begin{equation}
\Lambda=-3\tfrac{1+c_2+2c_3}{l^2}. 
\end{equation}
Note that $\Lambda$ can be positive, negative or
zero, depending on the value of the free parameters $c_{2}$ and $c_{3}$. The first term in the action (\ref{KomDejstvo}) (quadratic in curvature) is the Gauss-Bonnet term; it is topological in $D=4$ and does not affect the equations of motion.

\subsection{Dirac field}

One of the main advantages of the first-order formalism is its capacity to accommodate spinors in curved spacetime. 
Within the framework of AdS gauge theory of gravity, the Dirac spinor field $\psi$ transforms under the action of an $SO(2,3)$ gauge group as
\begin{equation} 
\delta_{\epsilon} \psi=i\epsilon\psi=\tfrac{i}{2}\epsilon^{AB}M_{AB}\psi,
\end{equation}
where $\epsilon^{AB}$ are antisymmetric gauge parameters. The $SO(2,3)$ covariant derivative is given by
\begin{align}
D_{\mu}\psi&=\partial_{\mu}\psi-\tfrac{i}{2}\omega_{\mu}^{AB}M_{AB}\psi\\ \nonumber
&=D^{L}_{\mu}\psi+\tfrac{i}{2l}e_\mu^{a}\gamma_{a}\psi,
\end{align} 
where $D^{L}$ stands for the Lorentz $SO(1,3)$ covariant derivative for spin-connection $\omega^{ab}_{\mu}$.

Using the auxiliary field $\phi=\phi^{A}\Gamma_{A}$, we can define a spinor action manifestly invariant under $SO(2,3)$ gauge transformations, 
\begin{align}\label{Dirac_before_SB}
\nonumber S_{\psi}=\frac{i}{12}\int 
&\de^{4}x\;\varepsilon^{\mu\nu\rho\sigma}\Big(\bar{\psi}D_{\mu}\phi D_{\nu}\phi 
D_{\rho}\phi D_{\sigma}\psi
\\
-&D_{\sigma}\bar{\psi}D_{\mu}\phi D_{\nu}\phi D_{\rho}\phi\psi\Big). 
\end{align}
Reduction of the $SO(2,3)$ gauge invariance down to the Lorentz $SO(1,3)$ invariance is done as before, and the action (\ref{Dirac_before_SB}) becomes   
\begin{align} \label{Dirac in curved space}
S_{\psi, g.f.}=i&\int\de^{4}x\sqrt{-g}\;\bar{\psi}\gamma^{\mu}\overleftrightarrow{D}^{L}_{\mu}\psi \nonumber\\
-\frac{2}{l}&\int 
\de^{4}x\sqrt{-g}\;\bar{\psi}\psi,
\end{align}
with $\gamma^{\mu}=e_{a}^{\mu}\gamma^{a}$. This is exactly the Dirac action in curved spacetime for spinors with ``mass'' $2/l$. Note, however, that in the presence of the cosmological constant, Minkowski space is not a solution of the field equations and therefore the definition of particle mass requires more attention. 

There are five additional, Yukawa-like terms that could supplement the original spinor action (\ref{Dirac_before_SB}). They differ only by the position of the auxiliary field $\phi$, for example
\begin{equation}
\bar{\psi}\De\phi\wedge\De\phi 
\wedge\De\phi\wedge\De\phi \phi\psi.    
\end{equation}
At the classical level, after the symmetry breaking, these terms only modify the fermionic AdS mass term. In particular, by tuning the coefficients, we can set the mass of the fermion to any given value. On the other hand, in the noncommutative setting, these Yukawa-like terms give some new effective interactions. 

\subsection{Yang-Mills field}

Introducing non-Abelian $SU(N)$ gauge field (similar consideration holds for the Abelian $U(1)$ case, that we will need later), $A_{\mu}=A_{\mu}^{I}T_{I}$, requires an upgrade of the original gauge group $SO(2,3)$ to $SO(2,3)\times SU(N)$. Generators $T_{I}$ of $SU(N)$ group are hermitian, traceless and they satisfy the commutation relations $[T_{I}, T_{J}]=if_{IJ}^{\;\;\;\;K}T_{K}$ with antisymmetric structure constants $f_{IJK}$. We use the normalization $Tr(T_{I}T_{J})=\delta_{IJ}$. $SU(N)$ group indices $I,J,\dots$ run from $1$ to $N^{2}-1$. The total gauge potential of $SO(2,3)\times SU(N)$ group is
\begin{equation}
\Omega_{\mu}=\frac{1}{2}\omega^{AB}_{\mu}M_{AB}\otimes\mathbb{1}+\mathbb{1}\otimes A_{\mu}^{I}T_{I}, \label{master-potential}
\end{equation}  
and the corresponding total field strength $\mathbb{F}_{\mu\nu}$ is the sum of the $SO(2,3)$ part $F_{\mu\nu}$ and the $SU(N)$ part $\mathcal{F}_{\mu\nu}$,
\begin{equation}
\mathbb{F}_{\mu\nu}=\frac{1}{2}F_{\mu\nu}^{AB}M_{AB}\otimes \mathbb{1}+\mathbb{1}\otimes \mathcal{F}_{\mu\nu}^{I}T_{I}, \label{master-field-strength}
\end{equation}
with the usual $\mathcal{F}^{I}_{\mu\nu}=\partial_{\mu}A^{I}_{\nu}-\partial_{\nu}A^{I}_{\mu}
+gf^{I}_{\;JK}A^{J}A^{K}$, where $g$ is the Yang-Mills coupling strength. 

An action for the Yang-Mills gauge field $A$ (the pure gravitational part will not contribute), invariant under $SO(2,3)\times SU(N)$ transformations, can be defined as
\begin{align}
S_{A}=&-\frac{1}{16l}\int \de^{4}x\;\varepsilon^{\mu\nu\rho\sigma}\tr\Big(f\mathbb{F}_{\mu\nu} \mathcal{D}_{\rho}\phi \mathcal{D}_{\sigma}\phi \phi \nonumber \\ 
&+\frac{i}{6}f^{2}\mathcal{D}_{\mu}\phi \mathcal{D}_{\nu}\phi \mathcal{D}_{\rho}\phi \mathcal{D}_{\sigma}\phi \phi \Big)+h.c. \label{ActionYM}
\end{align}
Note that now the covariant derivative $\mathcal{D}_\mu$ has also an $SU(N)$ component (though scalar $\phi$ is not charged under this symmetry). 
Action (\ref{ActionYM}) involves an additional auxiliary field $f$ defined by 
\begin{equation}
f=\frac{1}{2}f^{AB, I}M_{AB}\otimes T_{I}, \quad \delta_\epsilon f = i[\epsilon, f], \label{f} 
\end{equation}
where the gauge parameter is now 
\begin{equation}
\epsilon = \frac{1}{2}\epsilon^{AB}M_{AB}\otimes\mathbb{1}+\mathbb{1}\otimes\epsilon^{I}T_{I}. \end{equation}
The field $f$ transforms in the adjoint representation of $SO(2,3)$ and $SU(N)$ group. The role of this field is to produce the canonical kinetic term in curved spacetime for the $SU(N)$ gauge field in the absence of the Hodge dual operation, which otherwise cannot be defined without prior knowledge of the metric tensor. 

Again, by setting $\phi^{a}=0$ and $\phi^{5}=l$ in (\ref{ActionYM}) we break the $SO(2,3)$ gauge invariance and obtain
\begin{align}
S_{A}=&\frac{1}{2}\int \de^{4}x\;e\;  f^{ab,I}\mathcal{F}_{\mu\nu}^{I}e_{a}^{\mu} e_{b}^{\nu} \nonumber \\
+&\frac{1}{4}\int \de^{4}x\;e\;(f^{ab,I}f_{ab}^{I}+2f^{a5,I}f_{a5}^{I}). \label{ActionYM-afterSB}
\end{align}
where $e=\det(e^{a}_{\mu})$. Varying this action independently in components $f_{ab}^{I}$ and $f_{a5}^{I}$ of the auxiliary field, we get their equations of motion,
\begin{align}\label{YMEOMs}
f_{a5}^{I}=0,\;\;\;\;f_{ab}^{I}=-e_{a}^{\mu}e_{b}^{\nu}\mathcal{F}_{\mu\nu}^{I}.
\end{align}
Evaluating action (\ref{ActionYM-afterSB}) using (\ref{YMEOMs}) we eliminate the field $f$ and obtain
\begin{eqnarray}
S_{A}&=&-\frac{1}{4}\int \de^{4}x\;\sqrt{-g}\;g^{\mu\rho}g^{\nu\sigma}\mathcal{F}_{\mu\nu}^{I}\mathcal{F}_{\rho\sigma}^{I},
\end{eqnarray} 
which is exactly the canonical kinetic term for the Yang-Mills gauge field in curved spacetime.

The commutative fermionic action in the Yang-Mills theory after the symmetry breaking reads,  
\begin{align}
S_{\Psi}=&i\int \de^{4}x \sqrt{-g}\Big(i\bar{\Psi}\gamma^{\mu}\overleftrightarrow{D}^{L}_{\mu}\Psi
-\frac{2}{l}\bar{\Psi}\Psi \nn\\ &+g\bar{\Psi}\gamma^{\mu}A_{\mu}^{I}T_{I}\Psi\Big), \label{SPsi-afterSB}
\end{align}
where $\Psi$ is an $N$-tuple of Dirac spinors that transforms in the defining representation of the $SU(N)$ gauge group. 

\section{Noncommutative $SO(2,3)$ gravity}\label{gravitacija}

Now we consider a noncommutative extension of the $SO(2,3)$-invariant classical actions defined in previous sections. We give a short review of the Seiberg-Witten (SW) construction and present some main results obtained by analyzing NC corrections. 

\subsection{Seiberg-Witten map}

To set up a stage for NC gravity, we will briefly review the general structure of NC gauge field theories in the SW framework. Let $\mathfrak{g}$ be a Lie algebra with generators $\{T_{I}\}$. The basic elements of a classical (i.e. commutative) gauge field theory are the $\mathfrak{g}$-valued gauge field, $A_{\mu}=A_{\mu}^{I}T_{I}$, and the corresponding field strength, $F_{\mu\nu}=F_{\mu\nu}^{I}T_{I}$. Under an infinitesimal gauge transformations with a gauge parameter $\epsilon=\epsilon^{I}T_{I}$, they change as 
\begin{align}\label{varA}
    \delta_{\epsilon}A_{\mu}&=\partial_{\mu}\epsilon-i[A_{\mu},\epsilon],\\
    \delta_{\epsilon}F_{\mu}&=i[\epsilon,F_{\mu\nu}],\label{varF}
\end{align}
and the algebra of infinitesimal gauge transformations closes in the Lie algebra itself, 
\begin{equation}
[\delta_{\epsilon_{1}},\delta_{\epsilon_{2}}]=\delta_{i[\epsilon_{1},\epsilon_{2}]}.
\end{equation}
Covariant derivative acts on $F_{\m\nu}$ (or any field in the adjoint representation) as 
\begin{equation}
D_{\mu}F_{\rho\sigma}=\partial_{\mu}F_{\rho\sigma}-i[A_{\mu},F_{\rho\sigma}].
\end{equation}

Commutative gauge field theory actions can be promoted to their NC counterparts by introducing an NC star product. However, this seemingly simple step is not by itself enough to make a transition to a proper NC theory. To preserve the symmetry of the action, we also have to introduce NC fields (they will be denoted by a hat symbol) that change under NC gauge transformations in the same way as classical fields change under ordinary gauge transformations. For an NC gauge parameter $\widehat{\epsilon}$, infinitesimal NC variations of the NC gauge field $\widehat{A}_{\mu}$ and the NC field strength $\widehat{F}_{\mu\nu}$ are thus       
\begin{align}\label{NCvarA}
\widehat{\delta}_{\widehat{\epsilon}}\widehat{A}_{\mu}&=\partial_{\mu}\widehat{\epsilon}-i[\widehat{A}_{\mu}\ds\widehat{\epsilon}],\\
\widehat{\delta}_{\widehat{\epsilon}}\widehat{F}_{\mu\nu}&=i[\widehat{\epsilon}\ds\widehat{F}_{\mu\nu}]. \label{NCvarF}
\end{align}
which amounts to an NC deformation of (\ref{varA}) and (\ref{varF}). The resulting NC action is invariant under these deformed gauge transformations by construction. 

A commutator between two infinitesimal NC gauge transformations acts on the field $\widehat{F}$ as
\begin{align}
[\widehat{\delta}_{\widehat{\epsilon}_{1}},\widehat{\delta}_{\widehat{\epsilon}_{2}}]\widehat{F}_{\mu\nu}&=\widehat{\delta}_{i[\widehat{\epsilon}_{1}\ds\widehat{\epsilon}_{2}]}\widehat{F}_{\mu\nu}, 
\end{align}
with
\begin{align}\label{antikom}
[\widehat{\epsilon}_{1}\ds\widehat{\epsilon}_{2}]=\tfrac{1}{2}\big([\widehat{\epsilon}_{1}^{K}\ds\widehat{\epsilon}_{2}^{L}]&\{T_{K},T_{L}\} \nn\\
+&\{\widehat{\epsilon}_{1}^{K}\ds\widehat{\epsilon}_{2}^{L}\}[T_{K},T_{L}]\big).  
\end{align}
As anticommutator $\{T_{K},T_{L}\}$ appears in the above formula, infinitesimal NC gauge transformations do not generally close in the Lie algebra. One way to overcome this difficulty is to consider a larger, universal enveloping algebra (UEA) of the original Lie algebra, i.e., to assume that the NC gauge transformations parameter is UEA-valued, implying that the NC gauge field $\widehat{A}_\mu$ and NC field strength $\widehat{F}_{\mu\nu}$ are also UEA-valued. 

Unfortunately, as UEA is infinite-dimensional, this procedure seemingly introduces infinitely many new degrees of freedom. Though those new fields can be in principle used to tackle the problem of a dark matter, we find this introduction of fields physically unacceptable. Seiberg-Witten map enables us to express all the new degrees of freedom in terms of the classical ones \cite{SW, UEA}. In \cite{SW}, the main idea of this map was to relate NC fields and their commutative counterparts via perturbative in $\theta$ expansion, where coefficients of this expansion are determined in therms of classical fields, that transform under usual (commutative) gauge transformations.

This is achieved insisting that the NC field transformations are induced by the commutative ones
\begin{equation}
\widehat{\delta}_{\widehat{\epsilon}}\widehat{A}_\mu(A)=\widehat{A}_\mu(A+\delta_{\epsilon}A)-\widehat{A}_\mu(A),    
\end{equation}
where the NC gauge field $\widehat{A}$ is a function of the classical gauge field $A$ and $\widehat{\epsilon}=\widehat{\epsilon}(\epsilon,A)$. Using (\ref{varA}) and (\ref{NCvarA}) we can solve this differential equation perturbatively and derive the nonlinear SW map that represents NC fields $\widehat{\epsilon}$,  $\widehat{A}_\mu$ and $\widehat{F}_{\mu\nu}$ as a power series in $\theta$, with coefficients built out solely of classical fields. Up to first-order in $\theta$ we have 
\begin{align}
\widehat{\epsilon}&=\epsilon-\frac{1}{4}\theta^{\rho\sigma}\{A_{\rho},\partial_{\sigma}\epsilon\},\\
\widehat{A}_{\mu}&=A_{\mu}-\frac{1}{4}\theta^{\rho\sigma}\{A_{\rho},\partial_{\sigma}A_{\mu}+F_{\sigma\mu}\},\\
\widehat{F}_{\mu\nu}&=F_{\mu\nu}-\frac{1}{4}\theta^{\rho\sigma}\{A_{\rho},(\partial_{\sigma}+D_{\sigma})F_{\mu\nu}\} \nn\\ +&\frac{1}{2}\theta^{\rho\sigma}\{F_{\mu\rho},F_{\nu\sigma}\}.
\end{align}

We can use this mapping to expand NC gauge theory action in powers of $\theta$, where we are guaranteed that at each order action is (classically) gauge invariant. Note also that the reason for SW map was primarily the existence of anticommutator in (\ref{antikom}), that in general sits outside the Lie algebra. This also hints that there are situations (e.g. $U(N)$ algebras) where one does not have to go through this construction in order to talk about NC gauge theories, though in our work this is mandatory.

The NC generalizations of the classical AdS gauge gravity actions (\ref{KomDejstvo_S_1}), (\ref{KomDejstvo_S_2}) and (\ref{KomDejstvo_S_3}) are

\begin{align}\label{NCAction1} 
S^{\star}_{1}=\frac{ilc_{1}}{64\pi G_N}\int&{\rm d}^4x\; \varepsilon^{\mu\nu\rho\sigma}\\
&\times\tr\Big(\widehat{F}_{\mu\nu}\star \widehat{F}_{\rho\sigma}\star \widehat{\phi}\Big), \nonumber
\end{align}

\begin{align}\label{NCAction2} 
S^{\star}_{2}=\frac{c_{2}}{128 \pi G_{N}l}&\int \de^{4}x\; \varepsilon^{\mu \nu 
\rho \sigma} \\
\times&\tr\Big(\widehat F_{\mu \nu}\star D_{\rho}\widehat\phi\star 
D_{\sigma}\widehat\phi\star\widehat\phi\Big) + c.c., \nn
\end{align}
\begin{align}\label{NCAction3} 
S^{\star}_{3}=-\frac{ic_{3}}{128 \pi G_{N}l}&\int \mathrm{d}^{4} x\,
\varepsilon^{\mu\nu\rho\sigma} \\
\times\tr\Big(D_{\mu} 
\widehat\phi &\star D_{\nu} \widehat\phi\star
D_{\rho}\widehat{\phi} \star D_{\sigma}\widehat{\phi}\star \widehat{\phi}\Big).\nn
\end{align}
Note that at all points one should add $+h.c.$, in order to make NC action real. The SW expansions of the NC auxiliary field $\widehat{\phi}$ and its covariant derivative, up to first-order in $\theta$, are given by
\begin{align}
\widehat{\phi}&=\phi -\tfrac{1}{4}\theta^{\rho\sigma}
\{\omega_\rho,(\partial_\sigma + D_\sigma)\phi\}, \label{phi-NCexp}  \\
D_{\mu}\widehat{\phi}&=D_{\mu}\phi-\tfrac{1}{4}\theta^{\rho\sigma}\{\omega_{\rho},
(\partial_{\sigma}+D_{\sigma})
D_{\mu}\phi\} \nn\\
&+\tfrac{1}{2}\theta^{\rho\sigma}\{F_{\rho\mu},D_{\sigma}\phi\}\label{Dphi-NCexp}.
\end{align}

For computing first-order NC correction of a product of two NC fields we use the following general rule,
\begin{align}\left(\widehat{f}\star\widehat{g}\right)^{(1)}=&-\tfrac{1}{4}\theta^{\rho\sigma}\{A_\rho,(\partial_\sigma+D_\sigma)fg\} \nn\\
&+\tfrac{i}{2}\theta^{\rho\sigma} D_\rho f D_\sigma g \nn \\&+\text{cov}(\widehat{f}^{(1)})g+f\text{cov}(\widehat{g}^{(1)}), \label{rule-adj}\end{align}    
where $\text{cov}(\widehat{f}^{(1)})$ is the covariant part of $f$'s first-order NC correction, and likewise for $g$. Successive application of the expansion rule (\ref{rule-adj}) on NC actions (\ref{NCAction1}), (\ref{NCAction2}) and (\ref{NCAction3}) yields a perturbative expansion in powers of $\theta$. 
By construction, the expansion is invariant under $SO(2,3)$ gauge transformations order-by-order. After imposing the gauge fixing condition, the first-order NC correction vanishes while the second-order NC correction $S_{NC}^{(2)}$ can be found explicitly.\footnote{It is important to note that NC deformation does not commute with the gauge fixing. Therefore, one must first expand the NC action in powers of $\theta$ using the SW map and then apply the gauge fixing condition.}
The result is intricate, and we will not write the full expression here. We refer the reader to \cite{Us2, Us3}. However, one can still analyze the model in different regimes of parameters. If we are interested in the low-energy sector of the theory, we should keep only terms that have at most two derivatives on vierbeins. Therefore, we include only terms linear in curvature and linear and quadratic in torsion (we assumed that the spin-connection and the first-order derivatives of vierbeins are of the same order). The equations of motions are obtained by varying the action over vierbein and spin-connection, independently. If we consider only the class of NC solutions with classically vanishing torsion $T_{\mu\nu}^{a}=0$, in the low-energy limit, equations of motion for the vierbein and the spin-connection are, respectively
\begin{align}
& R_{\alpha\beta}^{cd}e_a^\alpha e^\beta_d e_c^\mu-\tfrac{1}{2}e^\mu_a R+\tfrac{3}{l^2}(1+c_2+2c_3)e^\mu_a = \tau_a^{\;\;\mu},\nn\\
& T_{ac}^{\ \ c}e_b^\mu-T_{bc}^{\ \ c}e_a^\mu-T_{ab}^{\ \ \mu} =
S_{ab}^{\ \
\mu}.\label{EoMT}
\end{align}
The effective energy-momentum tensor $\tau_a^{\ \mu}$ and the effective spin-tensor
$S_{ab}^{\ \ \mu}$ in equations (\ref{EoMT}) depend on $\theta^{\mu\nu}$ and we can conclude that noncommutativity acts as a source of curvature and torsion, i.e. spacetime can 
become curved and can develop a torsion due to NC
corrections. 

\subsection{NC Minkowski space}

In order to understand some consequences of spacetime noncommutativity, we consider the NC deformation of Minkowski space in the low-energy limit \cite{Us4}. Minkowski space is a vacuum solution of the Einstein field equations with $\Lambda=0$. Therefore, if we recall that $\Lambda=-3(1+c_2+2c_3)/l^2$,
we have to assume that $1+c_2+2c_3=0$ to eliminate the cosmological constant from the classical solution. Regarding NC correction as a small perturbation around flat Minkowski metric,
\begin{equation} 
g_{\mu\nu}=\eta_{\mu\nu}+l_{NC}^{-4}h_{\mu\nu},
\end{equation}
where $h_{\mu\nu}$ is quadratic in $\theta^{\mu\nu}\sim l_{NC}^{2}$, field equations reduce to 
\begin{align}\nonumber
&\tfrac{1}{2}(\partial_\sigma\partial^\nu h^{\sigma\mu}+\partial_\sigma\partial^\mu h^{\sigma\nu}-\partial^\mu\partial^\nu h -\Box 
h^{\mu\nu})\\& -\tfrac{1}{2}\eta^{\mu\nu}(\partial_\alpha\partial_\beta h^{\alpha\beta}-\Box h)\nn\\
&=l_{NC}^{4}\tfrac{11}{4l^6}\left(2\eta_{\alpha\beta}\theta^{\alpha\mu}\theta^{\beta\nu}+\tfrac{1}{2}g^{\mu\nu}\theta^{2}\right).
\label{NCMinkEoMmetric} 
\end{align}

The NC-deformed components of the metric tensor are given by 
\begin{align}
g^{00}&= 1 - \tfrac{11}{2l^6}\theta^{0m}\theta^{0n}x^m x^n-\tfrac{11}{8l^6}\theta^{\alpha\beta}\theta_{\alpha\beta}r^2,\nn\\
g^{0i} &=
-\tfrac{11}{
3l^6}\theta^{0m}\theta^{in}x^m 
x^n ,\nn\\ 
g^{ij}&= -\delta^{ij}-\tfrac{11}{6l^6}\theta^{im}\theta^{jn} 
x^mx^n \nn\\
&+\tfrac{11}{24l^6}\delta^{ij}\theta^{\alpha\beta}\theta_{\alpha\beta}r^2-\tfrac{11}{24l^6}
\theta^{\alpha\beta}
\theta_{\alpha\beta}x^i x^j. \label{NCMinkowskiMetric}
\end{align}
The Reimann tensor for this solution can be calculated easily and the scalar curvature of the NC Minkowski space turns out to be
$R=\tfrac{11}{l^{6}}\theta^{2}$, which is constant. Therefore, in the $SO(2,3)_{\star}$ model, there exists a non-trivial NC deformation of Minkowski space. A very 
interesting conclusion emerges: having the components of the Riemann tensor, the components of the metric
tensor can be represented as
\begin{align}
g_{00}&=1-R_{0m0n}x^mx^n,\nn\\ 
g_{0i}&=-\tfrac{2}{3}R_{0min}x^mx^n,\nn\\ g_{ij}&=-\delta_{ij}-\tfrac{1}{3}R_{imjn}x^mx^n.\label{FermiNCMinkowski}
\end{align}
As argued in \cite{Us3}, this suggests that the coordinates $x^\mu$ on which we impose the noncommutativity relations, are actually Fermi normal
coordinates. These are the inertial coordinates of a local observer moving along a geodesic.
This hints that one should impose a constant noncommutativity relations between coordinates $x^{\mu}$ precisely in those coordinates. In an arbitrary reference frame, the NC deformation is obtained by an appropriate coordinate transformation. 

This model was further investigated in \cite{Bailey:2018ifc}, where the relations with  Lorentz-violating actions were maid.

\section{NC Electrodynamics}

In this section, we will focus on a $U(1)$-invariant extension of AdS gauge theory of gravity and discuss some of its noncommutative aspects. The NC actions we are concerned with are the $\star$-product versions of the spinor action (\ref{Dirac_before_SB}), 
\begin{align}\label{DiracNC}
S^{\star}_{\psi}=&\frac{i}{12}\int 
\de^{4}x\;\varepsilon^{\mu\nu\rho\sigma}\nn\\
&\times\Big(\widehat{\bar{\psi}}\star\mathcal{D}_{\mu}\widehat{\phi} \star\mathcal{D}_{\nu}\widehat{\phi} 
\star\mathcal{D}_{\rho}\widehat{\phi}\star\mathcal{D}_{\sigma}\widehat{\psi} \nn\\
&-\mathcal{D}_{\sigma}\widehat{\bar{\psi}}\star\mathcal{D}_{\mu}\widehat{\phi}\star\mathcal{D}_{\nu}\widehat{\phi}\star\mathcal{D}_{\rho}\widehat{\phi}\star\widehat{\psi}\Big),
\end{align}
and the Abelian case of (\ref{ActionYM}),
\begin{align}\label{NCActionGauge}
S^{\star}_{A}&=-\frac{1}{16l}\int{\rm d}^{4}x\;\varepsilon^{\mu\nu\rho\sigma}\nn\\
&\times\tr\Big(\widehat{f}\star
\widehat{\mathbb{F}}_{\m\n}\star \mathcal{D}_\r\widehat{\phi}\star \mathcal{D}_\s\widehat{\phi}\star \widehat{\phi} \\
 &+ \frac{i}{3!}\widehat{f}\star\widehat{f}\star \mathcal{D}_\m\widehat{\phi}\star
\mathcal{D}_\n\widehat{\phi} \star \mathcal{D}_\r\widehat{\phi}\star \mathcal{D}_\s\widehat{\phi}\star \widehat{\phi}\Big)+ c.c. \nn
\end{align}
The NC Dirac spinor field $\widehat{\psi}$ changes under infinitesimal NC gauge transformations as
$\widehat{\delta}_{\widehat{\epsilon}} \widehat{\psi}= i \widehat{\epsilon} 
\star{\widehat
\psi}$, and likewise its covariant derivative. Their SW expansions are given by
\begin{align}
\widehat{\psi}&=\psi-\tfrac{1}{4}\theta^{\rho\sigma}\Omega_{\rho}(\partial_{\sigma}+\mathcal{D}_{\sigma})\psi, \label{psi-expansion} \\ 
\mathcal{D}_\mu\widehat{\psi} &= \mathcal{D}_\mu\psi-
\tfrac{1}{4}\theta^{\rho\sigma}\Omega_\rho(\partial_\sigma 
+\mathcal{D}_\sigma)\mathcal{D}_\mu\psi \nn\\
&+\tfrac{1}{2}\theta^{\rho\sigma}\mathbb{F}_{\rho\mu}\mathcal{D}_\sigma\psi,
\end{align}

The NC action (\ref{NCActionGauge}) involves an NC auxiliary field $\widehat{f}$, classically defined in (\ref{f}), that transforms in the adjoint representation of the $SO(2,3)_{\star}\times U(1)_{\star}$ gauge group, i.e. $\widehat{\delta}_{\widehat{\epsilon}}\widehat f = i[\widehat{\epsilon}\ds\widehat {f}]$. 
The transformation laws of the NC fields ensure the invariance of actions (\ref{DiracNC}) and (\ref{NCActionGauge}) under $SO(2,3)_{\star}\times U(1)_{\star}$ NC gauge transformations.

After the symmetry breaking, this NC model amounts to an effective theory of interacting Dirac fermions in curved spacetime that involves many new types of coupling due to spacetime noncommutativity. Moreover, there are some residual effects of spacetime noncommmutativity that survive even the flat spacetime limit. Therefore, from now on we work in Minkowski space \footnote{As we saw in the previous section, Minkowski space can be considered as a classical solution if we exclude the cosmological constant by a suitable choice of coefficients. Also, Minkowski space receives NC corrections at the second-order in $\theta$, and we can therefore use only the classical Minkowski metric when working at the first NC order.}.  

The action for NC electrodynamics in Minkowski space, up to first-order in $\theta$, is given by
\begin{align}\label{SQED-flat}
&S^{\star}_{flat}= \int {\rm d}^{4}x\;\bar{\psi}(i\slashed{\mathcal{D}} - m)\psi-\frac{1}{4}\int \de^{4}x\;\mathcal{F}_{\m\n}\mathcal{F}^{\m\n}\nn\\
&+\theta^{\alpha\b}\int {\rm d}^{4}x\;\Bigg(\frac{1}{2}\mathcal{F}_{\alpha\m}\mathcal{F}_{\b\n}\mathcal{F}^{\m\n}-\frac{1}{8}\mathcal{F}_{\alpha\b}\mathcal{F}^{\m\n}\mathcal{F}_{\m\n} 
\Bigg) \nn\\
&+\theta^{\alpha\b}\int {\rm d}^{4}x\;\bar{\psi}\Bigg(-\frac{1}{2l}
\sigma_{\alpha}^{\;\;\s}\mathcal{D}_{\b}\mathcal{D}_{\s}+\frac{3i}{4}\mathcal{F}_{\alpha\b}\slashed{\mathcal{D}}\nn\\
&+\frac{7i}{24l^{2}}\varepsilon_{\alpha\b}^{\;\;\;\;\r\s}
\gamma_{\r}\gamma_{5}\mathcal{D}_{\s}-\frac{i}{2}\mathcal{F}_{\alpha\m}\gamma^\m\mathcal{D}_\b  \\
&-\left(\frac{m}{4l^{2}}+\frac{1}{6l^{3}}\right)\sigma_{\alpha\beta} 
-\left(\frac{3m}{4}-\frac{1}{4l}\right)\mathcal{F}_{\alpha\b}\Bigg)\psi, \nn
\end{align}
where $\mathcal{D}_{\m}=\partial_{\m}-iA_{\m}$ is now a flat spacetime covariant derivative. The $\theta$-dependent terms represent new types of interactions that introduce some new effects such as a modification of the Landau levels of an electron. Actions describing NC Electrodynamics are also studied in \cite{MinNCED, NonRenNCQED, NCED}. 

\subsection{Electron in a background magnetic field}

By varying NC action (\ref{SQED-flat}) over $\bar{\psi}$, we obtain a $\theta$-deformed Dirac equation for an electron (we set $q=-1$) coupled to electromagnetic field,
\begin{equation}
\left(i\slashed{\partial}-m+\slashed{A} 
+\theta^{\alpha\b}\mathcal{M}_{\alpha\b}\right)\psi=0, \label{EM-Dirac}
\end{equation}
with the NC correction
\begin{align}
\theta^{\alpha\b}&\mathcal{M}_{\alpha\b}=\theta^{\alpha\b}\Bigg\{-\frac{1}{2l}\sigma_\alpha^{\;\;\s}\mathcal{D}_\b\mathcal{D}_\s+\frac{3i}{4}\mathcal{F}_{\alpha\b}\slashed{\mathcal{D}} \nn\\
&+\frac{7i}{24l^{2}}\epsilon_{\alpha\b}^{\;\;\;\;\r\s}\gamma_\r\gamma_5 \mathcal{D}_\s-\frac{i}{2}\mathcal{F}_{\alpha\m}\gamma^\m\mathcal{D}_\b \\
&- \left(\frac{m}{4l^{2}}+\frac{1}{6l^{3}}\right)\sigma_{\alpha\b} -\left(\frac{3m}{4}-\frac{1}{4l}\right)\mathcal{F}_{\alpha\b}\Bigg\}. \label{thetaM} \nn
\end{align}
We can now analyze a special case of an electron propagating in constant magnetic field $\boldsymbol{B}=B\boldsymbol{e}_{z}$. Accordingly, we choose $A_{\m}=(0, By, 0, 0)$. An appropriate ansatz for (\ref{EM-Dirac}) is  
\begin{equation}
\psi=\begin{pmatrix}
\varphi(y)   \\[6pt]
\chi(y)       
     \end{pmatrix}e^{-iEt+ip_{x}x+ip_{z}z}. \label{ansatz}
\end{equation}
Spinor components and the energy function are  represented as perturbation series in powers of $\theta$, 
\begin{align}
\varphi &=\varphi^{(0)}+\varphi^{(1)}+\mathcal{O}(\theta^{2}), \\
\chi &=\chi^{(0)}+\chi^{(1)}+\mathcal{O}(\theta^{2}), \\
E&=E^{(0)}+E^{(1)}+\mathcal{O}(\theta^{2}).
\end{align}
The unperturbed energy levels of an electron in constant magnetic field (Landau levels) are
\begin{equation} 
E_{n,s}^{(0)}=\sqrt{p_{z}^{2}+m^{2}+(2n+s+1)B}, \label{E-undeformed}
\end{equation}
where $n=0,1,2,\dots$ is the quantum number that determines the Landau level, while $s=\pm 1$ is the eigenvalue of the Pauli matrix $\sigma_3$. 

Inserting the ansatz (\ref{ansatz}) in the NC Dirac equation (\ref{EM-Dirac}) gives us the first-order NC correction to the Landau levels,
\begin{equation}
E^{(1)}_{n,s} = -\frac{\theta^{\alpha\b}\int \de y\;\bar\psi^{(0)}_{n,s}{\cal M}_{\alpha\b}\psi^{(0)}_{n,s}}{\int \de y\;\bar\psi^{(0)}_{n,s}\g^{0}\psi^{(0)}_{n,s}}. \label{EnergyCorrectionExpl}
\end{equation}
In particular, for $\theta^{12}=-\theta^{21}=\theta\neq 0$ and all the other components $\theta^{\alpha\b}$ equal to zero, we find
\begin{align}
E^{(1)}_{n,s}=-\frac{\theta s}{E^{(0)}_{n,s}}&\left(\frac{m^{2}}{12l^{2}}-\frac{m}{3l^{3}}\right)\nn\\
\times\Bigg(1+ &\frac{B}{(E^{(0)}_{n,s}+m)}(2n+s+1)\Bigg) \nn  \\
+& \frac{\theta B^{2}}{2E^{(0)}_{n,s}}(2n+s+1). \label{NC-energy-correction}
\end{align}
If we take $B=0$, we see that NC energy levels depend on $s=\pm 1$ and so the NC background causes Zeeman-like splitting of the undeformed energy levels. We might say that NC geometry acts as a birefringent medium. 

Consider now an electron confined to a plane, i.e. $p_{z}=0$, and let us ignore the spin. In the non-relativistic limit, $B\ll m^{2}$, classical (undeformed) Landau levels can be expanded in powers of $B/m^{2}$, 
\begin{align} \label{Landau_nonrel}
E_{n,s}^{(0)}=m\Big[1&+\frac{(2n+1)B}{2m^{2}}\\
&-\frac{(2n+1)^{2}B^{2}}{8m^{4}}
+\mathcal{O}((B/m^{2})^{3})\Big].\nn
\end{align}
Likewise, the non-relativistic NC energy levels reduce to
\begin{align}
E_{n,s}= m\bigg[1 &+ \frac{(2n+1)B_{eff}}{2m^{2}}\nn \\
&-\frac{(2n+1)^{2}B_{eff}^{2}}{8m^{4}}+\mathcal{O}(\theta^{2})\bigg], \label{NC-energylevels-B}
\end{align}
where we introduced $B_{eff}=(B+\theta B^{2})$ as an effective magnetic field. Comparing (\ref{NC-energylevels-B}) with the classcial result (\ref{Landau_nonrel}), we see that the only effect of spacetime noncommutaivity is to modify the value of the background magnetic field. This result is in the spirit of what is known in string theory. Namely, in \cite{SW} it is argued that the endpoint coordinates of an open string constrained to a D-brane in the presence of a constant Kalb-Ramond B-field satisfy constant noncommutativity algebra. 

From the energy function (\ref{NC-energylevels-B}) we can derive the induced magnetic moment of an electron in the $n^{th}$ Landau level, for weak magnetic field,
\begin{align}
\mu_{n,s}&=-\frac{\partial E_{n,s}}{\partial B}=-\mu_{B}(2n+s+1)(1+\theta B), \label{NCMgMoment}
\end{align}
where $\mu_{B}=\tfrac{e\hslash}{2mc}$ is the Bohr magneton. We immediately recognize $-(2n+1)\mu_{B}$ as the diamagnetic moment of an electron and $-s\mu_{B}$ as the spin magnetic moment. The $\theta B$-term is another potentially observable phenomenological prediction. Canonical ($\theta$-constant) deformation od relativistic Landau levels has been studied in \cite{NCLLl} and for other types of NC spacetimes in \cite{NCLL2, NCLL3}. Using the results of this section, one could hope to obtain some constraints on the noncommutaivity parameter from condensed matter experiments. 

\section{Kaluza-Klein reduction from 5D Chern-Simons gauge theory}

After setting up the AdS gauge theory of gravity (and its NC extension), a question remains whether there is some additional motivation for focusing on this particular formulation of gravity. In this section, we discuss how the gauge-invariant topological gravity Lagrangian $\tr(F\wedge F \phi)$ may appear as an effective low-energy model of four-dimensional gravity. In particular, we provide an explanation for the origin of the auxiliary scalar field $\phi^A$. 

As we have seen, pure gravity (without matter) does not have a first-order correction in $\theta$. There is, however, one remarkable exception to this rule. It was shown in \cite{Polychronakos, Leo} that NC Chern-Simons (CS) theories have, generally, corrections to all orders in $\theta$, as long as we work in $D>3$. However, the CS Lagrangians are defined only for odd-dimensional manifolds, and therefore there are no obvious immediate phenomenological consequences of this fact. 

On the other hand, extra dimensions have been considered in physics for around a century. Soon after the development of GR, it was suggested by Kaluza and Klein that one should consider a pure gravity theory in five dimensions and reduce it to four dimensions by compactifying one spatial dimension \cite{Kaluza, Klein} - the method now known as the Kaluza-Klein (KK) dimensional reduction. This procedure yields a four-dimensional theory of gravity coupled to $U(1)$ gauge field, together with a scalar field (usually called the radion). With this in mind, one can dimensionally reduce $D=5$ CS gravity to obtain an effective four-dimensional theory of gravity (assuming that the compactification radius is small enough to not interfere with the observed physics and ignoring higher KK modes). For classical (commutative) $(2n-1)$-dimensional CS gravity, this was done in \cite{Chamseddine_topo} where it was shown that this procedure leads to a $(2n)$-dimensional topological gravity theory of the kind we discussed so far. The method of KK reduction can also be applied to an NC CS theory in $D=5$, yielding an effective $D=4$ theory of gravity with new $\theta$-dependent low-energy couplings. To the first-order in $\theta$, this is done in \cite{Dordevic:2022ruk}; here we comment on the results obtained therein.

Since in this section we are only interested in gravity, we will change our metric signature convention to ``mostly positive'' ($\eta_{ab}=(-,+,+,+)$). Also, Chern-Simons Lagrangians are written more elegantly in terms of differential forms (as one does not have to introduce a metric tensor to write them). We will use both component and form notation in what follows, as per convenience.  

Firstly, the CS Lagrangians in $D=2n-1$ dimensions are built from the top form $L_{CS}$ that satisfies $\de L_{CS}=const\cdot\tr (F^n)$, where $F$ is the curvature $2$-form of the relevant Lie algebra-valued gauge connection. The trace operator generally represents an invariant symmetric polynomial on the Lie algebra. Since we will use an explicit representation for the generators, $\tr$ stands for the ordinary trace of matrices. We consider $D=5$ spacetime ($n=3$).
In order to have a gravity interpretation, we use a conformal gauge group $SO(4,2)$, with the gauge connection given by 
\begin{equation}
A_{\hat{\mu}}=\frac{1}{2}\Omega_{\hat{\mu}}^{AB}J_{AB}+\frac{1}{l}E_{\hat{\mu}}^AP_A.
\end{equation}
A hat over a spacetime index indicates that it can take values from zero to four.
Inserting this expression into our definition of $L_{CS}$, and choosing an appropriate constant in the definition, one obtains the $D=5$ CS action,
\begin{align}\label{CS_first}\nonumber
S_{CS}=-&\frac{k}{8}\int\de^5x\; \varepsilon_{ABCDE}\varepsilon^{\hat{\mu}\hat{\nu}\hat{\rho}\hat{\sigma}\hat{\lambda}}\nn\\
\times\Bigg(&\frac{1}{4l}R_{\hat{\mu}\hat{\nu}}^{AB}R_{\hat{\rho}\hat{\sigma}}^{CD}E_{\hat{\lambda}}^{E} 
+\frac{1}{3l^{3}}R_{\hat{\mu}\hat{\nu}}^{AB}E_{\hat{\rho}}^{C}E_{\hat{\sigma}}^{D}E_{\hat{\lambda}}^{E}\nn\\
+&\frac{1}{5l^{5}}E_{\hat{\mu}}^{A}E_{\hat{\nu}}^{B}E_{\hat{\rho}}^{C}E_{\hat{\sigma}}^{D}E_{\hat{\lambda}}^{E}\Bigg). 
\end{align}
It can be written in a more familiar form as
\begin{align}
S_{CS}^{(5)}=&\frac{1}{16\pi G^{(5)}}\int\de^{5}x\sqrt{-g}\Big[R-2\Lambda \\
&+\frac{l^{2}}{4}\left(R^{2}-4R^{\hat{\mu}\hat{\nu}}R_{\hat{\nu}\hat{\mu}}+R^{\hat{\mu}\hat{\nu}\hat{\rho}\hat{\sigma}}R_{\hat{\rho}\hat{\sigma}\hat{\mu}\hat{\nu}}\right)\Big], \nonumber 
\end{align}
where we put $k=\frac{l^3}{8\pi G^{(5)}}$, and $G^{(5)}$ is the $D=5$ gravitational constant.
It must be noted, however, that here the underlying geometric structure is more general than in $D=4$ GR, as nonzero torsion is allowed by the equations of motion. 

The star-product formalism can be extended to include higher differential forms, enabling us to express all formulas in a coordinate-independent fashion. The star-wedge product between two arbitrary forms $\tau_{p}$ and $\tau_{q}$ is defined by
\begin{align}
\tau_{p}\wedge_{\star}\tau_{q}&=\sum_{n=0}^{\infty}\frac{1}{n!}\Big(\frac{i}{2}\Big)^n\theta^{I_1 J_1}\dots\theta^{I_nJ_n}\nn\\
&\times(\ell_{I_1}\dots\ell_{I_n}\tau_{p})\wedge(\ell_{J_1}\dots\ell_{J_n}\tau_{q}),\end{align}
where $\ell_I$ stands for the Lie derivative along a particular vector field that belongs to a set of mutually commuting vector fields $\{X_I\}$. These fields define the NC structure of spacetime and allow us to have a manifestly diffeomorphism-invariant theory. However, they also make an NC theory emphatically background-dependent. By identifying $\{X_{I}\}$ with some coordinate basis vector fields, one effectively chooses in which coordinate system will $\theta$-constant noncommutativity be realized. There is no definite criteria coming from physics of how to select these vector fields and whether there is a preferred coordinate systems in an NC field theory. When the
star-product is obtained via a set of mutually commuting vector fields (i.e. via an abelian
twist) one can construct order-by-order solutions to the SW map \cite{PLGR-fer2, Leo}. 

First-order NC correction to the $D=5$ CS action is computed in \cite{Leo} and it is given by
\begin{align}
&S_{CS,\theta}=\frac{k\theta^{IJ}}{12}\int
\nn\\
\bigg(&F^{AB}(F_{I})_{BC}(\De F_{J})^{C}_{\;\;A}+\frac{1}{l^{2}}F^{AB}(F_{I})_{BC}(T_{J})^{C}E_{A} \nonumber\\ +&\frac{1}{l^{2}}F^{AB}(T_{I})_{B}(\De T_{J})_{A}+\frac{2}{l^{2}}F^{AB}(T_{I})_{B}(F_{J})_{AC}E^{C}\nonumber\\
+&\frac{1}{l^{2}}T^{A}
(\De T_{I})^{B}(F_{J})_{BA}
+\frac{1}{l^{2}}T_{A}(F_{I})^{AB}(F_{J})_{BC}E^{C}
\nonumber\\  
+&\frac{1}{l^{2}}T^{A}(T_{I})^{B}(\De F_{J})_{BA}+\frac{2}{l^{4}}T_{A}(T_{I})_{B}(T_{J})^{[B}E^{A]}
\bigg).\nonumber
\end{align}
Index $I$ in $F_{I}$ means that the curvature is contracted with the vector field $X_I$. 
Upon dimensional reduction, followed by the standard truncation of fields and symmetry breaking, one obtains an effective low-energy model of four-dimensional gravity,
\begin{align}\label{celoS}
&S_{red,NC}=\frac{(2\pi R)k}{8l^{3}}\int\nn\\
&\varepsilon_{abcd}\bigg(l^{2}R^{ab}R^{cd}+2R^{ab}e^{c}e^{d}+\frac{1}{l^{2}}e^{a}e^{b}e^{c}e^{d}
\bigg) \nonumber\\
+&\frac{(2\pi R) k}{12} \theta^{I4}\int\bigg(\frac{2}{l^{4}}R^{ab}T_{a}(e_{I})_{b}-\frac{4}{l^{4}}T^{a}(R_{I})_{ab}e^{b}\nn\\
+&\frac{2}{l^{4}}R^{ab}(T_{I})_{a}e_{b}+\frac{6}{l^{6}}T^{a}e_{a}(e_{I})^{b}e_{b}\bigg). 
\end{align}
There are few interesting facts to be noted. First, only terms involving $\theta^{I4}$ survived, which means that noncommutativity between the compactified dimension and the remaining ones is what gives the NC correction at first-order in $\theta$. In fact, we do not have to assume NC relations between the extended dimensions themselves. The existence of the NC compactified dimension alone gives us an effective NC gravity in four dimensions. 
Moreover, this result is consistent with the fact that in four-dimensional gravity (without the extra KK dimension) the lowest order NC correction is of the second-order in $\theta$ \cite{PLM-13, Us2, Us3}. More generally,  it is proven in \cite{UlasSaka:2007ue} that KK reduction and NC deformation operations are interchangeable if we assume that only non-compactified coordinates fail to commute mutually. Therefore, an NC compactified dimension is essential for having a non-trivial
first-order NC effects in the KK-reduced theory.

By varying action (\ref{celoS}), one obtains equations of motions for the vierbein and the spin-connection. Before we proceed, we should make some comments on this procedure. The radius of compactification should be connected with $\sqrt{\theta}$, by a suitable analysis of dimensions. Furthermore, we cannot hope to solve the equations of motion for the most generic vector fields $X_I$, nor can we obtain those fields directly from our theory. We will therefore use a specific choice of the vector fields, and for the sake of simplicity, we will choose the most obvious one. 

As classical equations of motions are the usual Einstein equations with the negative cosmological constant $\Lambda=-3/l^{2}$, our starting point in the perturbative analysis will be some solution to these equations. It is easy to prove that, at first-order in $\theta$, the AdS spacetime remains the solution of the equations of motion, i.e. it does not receive NC corrections (in AdS spacetime, constant $l$ is the AdS radius, as promised). Another interesting solution to consider is the AdS-Schwarzschild black hole, with metric
\begin{equation}\label{AdSBH}
    \de s^2=-f^2(r)\de t^2+\frac{1}{f^2(r)}\de r^2 + r^2\de \Omega^2,
\end{equation}
with $f^2(r)\equiv \left(1-\frac{2m}{r}+\frac{r^2}{l^2}\right)$, where $m$ is proportional to the black hole's mass \cite{Socolovsky:2017nff}. At this point, we assume that only two mutually commuting vector fields $X_I$ are $\partial_r$ and $\partial_4$.
This classical torsionless geometry develops first-order NC correction in spin-connection while tetrads remain the same. The NC correction results in a non-vanishing torsion and non-zero Pontryagin density $R^{ab}R_{ab}$, consequently introducing a chiral gravitational anomaly. Here we provide only the final results from \cite{Dordevic:2022ruk}. The NC corrections to torsion components are 
\begin{align}
    &\Tilde{T}^3_{02}=-\frac{m\theta^{14}}{2lr^2},\nonumber\\
    &\Tilde{T}_{03}^2=\frac{m\theta^{14}}{2l}\frac{\sin \theta}{r^2},\nonumber\\
    &\Tilde{T}_{23}^0=-\frac{m\theta^{14}}{l}\frac{\sin \theta}{rf(r)}.
\end{align}
This information can be used to deduce the NC spin-connection and, finally, the first-order NC corrections to the curvature tensor, 
\begin{align}
\tilde{R}_{23}^{01}&=-\frac{m\theta^{14}}{l}\frac{\sin\theta}{r^{2}},\nonumber\\
\tilde{R}_{13}^{02}&=-\frac{m\theta^{14}}{l}\left[\frac{f'(r)}{2r^{2}}+\frac{f(r)}{r^{3}}\right]\frac{\sin\theta}{f^{2}(r)}=-\sin\theta\tilde{R}_{12}^{03},\nonumber\\
\tilde{R}_{03}^{12}&=\frac{m\theta^{14}}{l}\left[\frac{f'(r)}{2r^{2}}-\frac{f(r)}{r^{3}}\right]\sin\theta=-\sin\theta\tilde{R}_{02}^{13},\nonumber\\
\tilde{R}_{01}^{23}&=\frac{3m\theta^{14}}{lr^{4}}.
\end{align} 
One can now straightforwardly compute the Pontryagin density and find the chiral gravitational anomaly,
\begin{equation}
    \partial_\mu (\sqrt{-g}j_5^\mu)=\frac{m^2\theta^{14}}{2\pi^2lr^5}\sin \theta.
\end{equation}
Here, $j_5^{\mu}$ stands for the axial current of a massless (commutative) fermion coupled to a fixed NC background. 
This expression is valid for all $r>r_h$ including large distances and small cosmological constant. An important thing to note is that this anomaly contribution vanishes in the strict limit of zero cosmological constant.

\section{Outlook}

Finally, let us mention some open problems that require further consideration. First of all, the necessity of having a multiplet of scalar fields in the four-dimensional (even-dimensional, in general) topological gauge theory of gravity raises the question of how these scalars should be treated and interpreted. In this review, we proposed one plausible explanation by which the scalar fields naturally appear by performing the Kaluza-Klein reduction of the $D=5$ Chern-Simons action. However, there are other possible ways these fields could be introduced, and this affects the NC extension of the classical theory. Another problem is the impossibility of constructing an $SO(2,3)$-invariant kinetic action for a scalar field with non-vanishing first-order NC correction.
Furthermore, the result obtained in the last section concerning the chiral gravitational anomaly is an interesting aspect of quantum field theory on an NC background. However, one might wonder whether the result persists upon including other terms coming from Kaluza-Klein reduction, including the massive modes. On a more general side, in order to have a diffeomorphism-invariant NC theory, we have to introduce an additional structure in terms of a set of commuting vector fields. However, there are no obvious physical criteria for making an appropriate choice. Adapting the vector fields to a particular system of coordinates (the one in which the $\theta$-constant noncommutativity will be realized) results in a breakdown of diffeomorphism-invariance. To have a complete, background-independent NC theory with no preferred coordinate system, one has to understand how these vector fields emerge in an NC theory.   

\bmhead{Acknowledgments} M.D.\'{C}. thanks the editors Konstantinos Anagnostopoulos, Peter Schupp
and George Zoupanos for the invitation to contribute to this special issue of ''Noncommutativity and Physics''.
The authors acknowledge funding provided by the Faculty of Physics, University of Belgrade, through the grant by the Ministry of Education, Science, and Technological Development of the Republic of Serbia (number 451-03-68/2022-14/200162).



\end{document}